\documentclass[preprint,showpacs,showkeys,preprintnumbers,amsmath,amssymb]{revtex4}
\usepackage[]{graphicx}
\usepackage{amsmath}

\begin{document}

\title{On the discrete Peyrard-Bishop model of DNA: stationary solutions and stability}

\author{Sara Cuenda}
\email{scuenda@math.uc3m.es}

\author{Angel S\'anchez$^{*,}$}
\email{anxo@math.uc3m.es}

\affiliation{%
$^*$Grupo Interdisciplinar de Sistemas Complejos (GISC) and
Departamento de Matem\'aticas\\
Universidad Carlos III de Madrid, Avenida de la Universidad 30, 28911
Legan\'es, Madrid, Spain}

\homepage{http://gisc.uc3m.es}

\affiliation{%
$^{\dag}$Instituto de Biocomputaci\'on y  F\'{\i}sica de
Sistemas Complejos (BIFI),\\
Facultad de Ciencias, Universidad de Zaragoza, 50009 Zaragoza, Spain
}%

\date{\today}

\begin{abstract}
As a first step in the search of an analytical study of mechanical
denaturation of DNA in terms of the sequence, we study stable,
stationary solutions in the discrete, finite and homogeneous
Peyrard-Bishop DNA model. We find and classify all the stationary
solutions of the model, as well as analytic approximations of
them, both in the continuum and in the discrete limits. Our
results explain the structure of the solutions reported by
Theodorakopoulos {\em et al.} [Phys.\ Rev.\ Lett.\ {\bf 93},
258101 (2004)] and provide a way to proceed to the analysis of the
generalized version of the model incorporating the genetic
information.
\end{abstract}



\maketitle

\noindent{\bf LEAD PARAGRAPH}\\

{\bf DNA, the molecule that constitutes the basis of the genetic
code, is of utmost importance. In particular, its mechanical
properties are crucial as opening the double helix structure of
DNA is needed to read the genetic code and for replication of the
molecule for reproduction. The complete separation of the double
helix is called replication, and can be achieved by heating or
mechanically, by pulling the two strands of the molecule apart. We
here address the mathematical description of mechanical
denaturation in terms of a simple model. We determine and classify
the solutions of the model equations and study their stability
properties. We also provide an approximate but very accurate way
to deal analytically with those solutions. Beyond mechanical
features, our results are relevant for studies of the
thermodynamic properties of the DNA chain, and may have genomic
applications, in so far as mechanical denaturation experiments
that give information about DNA composition can be modelled by our
model and solutions.}

\newpage

\section{Introduction}
Nonlinear models appear in many fields of science since the
pioneering discoveries, almost 50 years ago, of Fermi, Pasta and
Ulam \cite{fpu}. This work, in the field of physics, has led many
scientist to use nonlinear models in the study of complex systems
\cite{scott} in other subjects. Nonlinear models entered into DNA
physics with Englander and co-workers \cite{englander} (see
\cite{yaku} for a review on nonlinear models of DNA), in 1980,
when they modeled the dynamics of DNA with a sine-Gordon equation.
Since then, a lot of work has been devoted to nonlinear
excitations in DNA, both from the dynamics and the statistical
mechanics points of view. Among this body of work, a particularly
succesful model is the Peyrard-Bishop (PB) one \cite{pb,DPB}, that
will be our starting point in this paper.

One problem of special interest in the framework of DNA was the
thermal denaturation transition, which takes place at temperatures
around 90$^\circ$C, when the two strands of the DNA molecule
separate. On the other hand, mechanical denaturation, that occurs
when one of the strands of the molecule is separated from the
other by pulling it in single molecule experiments, was achieved
in the last few years \cite{Nature}. In order to model these
phenomena, most of the research done so far refers to {\em
homopolymers}, i.e., homogeneous DNA molecules consisting entirely
of A-T or C-G base pairs. When the issue under discussion is
genomics, or gene identification, which is very much related to
the above mentioned problems, models of {\em heteropolymers} are
required: The distribution of A-T and C-G base pairs follows
non-uniform, nonhomogeneous sequences obtained from genome
analysis. The heterogeneous PB model is also being used for
identifying relevant sites, such as promoters
\cite{LosAlamos,nuevomichel} in viral sequences, and also for
analyzing the thermal denaturation process \cite{saul}.

The main motivation of this work is the study of the effects of
the sequence heterogeneity on the dynamics of the mechanical
denaturation process. We began that research program by analyzing
the Englander (basically, the sine-Gordon equation \cite{scott})
model. The results we obtained  \cite{sara1, sara2} showed that
the Englander model was much too simple to reproduce the phenomena
observed in experiments, and therefore we decided to focus on the
PB model (see \cite{nonlinear} for an authoritative review about
this model). In that context, our immediate aim was to obtain a
tool in this model similar the effective potential proposed for
the Englander model by Salerno and Kivshar \cite{Salerno,
Salerno2, Salkiv} in order to study the relation between the
dynamics of these excitations and gene identification. To that
end, it is necessary to obtain stationary states of the
homogeneous model. Those were available for the continuous version
of the PB model, but, in fact, DNA is quite a discrete system, and
the discretization parameter depends on experimental measurements
used as parameters in the model. For instance, in the PB model,
the dimensionless parameter that defines the effective
discretization of the system can go from $R=10.1$ (see next
section for a definition of $R$), used in \cite{nonlinear, pbdis},
to $R\simeq 75$ used in \cite{campa}, or even to $R=100$ in
\cite{qasmi}. In all cases, these $R$ values correspond to systems
that are far from the continuum limit. Therefore, as a first step
towards our chief goal of understand sequence effects on
denaturation, our immediate purpose is to study stable, stationary
states in the discrete PB model, with a special focus on their
dependence on the effective discretization.

In this paper, we aim to finding stationary solutions of the PB
model and their corresponding stability conditions. These issues
are addressed in Secs.\ \ref{sec:model}. Subsequently, we discuss
the validity of the continuum limit and the domain wall
approximation in Sec.\ \ref{sec:cont}, while in the main part of
the paper, Sec.\ \ref{sec:discrete}, we propose analytical
approximations for the discrete case and compare our results with
the ones obtained in \cite{pbdis}. Finally, Sec.\
\ref{sec:conclusions} concludes the paper by summarizing our main
results and their possible implications.

\section{Discrete solutions and stability}\label{sec:model}

In the following we will use the dimensionless PB model, defined
by the hamiltonian
\begin{equation}\label{ham}
H=\sum_{n=0}^{N-1}\left\{\frac{1}{2}\dot Y_n^2 +\frac{1}{2R}(Y_{n+1}-Y_n)^2
+V(Y_n) \right\},
\end{equation}
where $V(Y)=(1-e^{-Y})^2$ is the Morse potential, that stands for
the atraction between the two bases of a base pair, and $R$ is a
positive, dimensionless constant that refers to the intensity of
the coupling of two consecutive bases. This constant plays the
role of an effective discretization, $a=\sqrt{R}$, so that $R\gg
1$ stands for a large discretization and $R\ll 1$ is the
continuous limit.

Static solutions of hamiltonian (\ref{ham}) must satisfy $\partial
H/\partial Y_n=0$, which turns out to be the recurrence relation
\begin{equation}\label{rec1}
Y_{n+1}=2Y_n-Y_{n-1}+RV'(Y_n),
\end{equation}
for $n=1,2,\ldots,N$. These solutions are uniquely defined by the
initial condition $\{Y_0,Y_1\}$. If we restrict ourselves to
solutions with $Y_0=0$, which we can do without loss of
generality, then each $Y_n$ will only depend on the value $Y_1=y$,
so that Eq.\ (\ref{rec1}) can be rewritten in terms of $y$
introducing the notation $Y_n(y)$ instead of $Y_n$. From now on we
will discuss the behavior of the solutions $Y_n(y)$ as a function
of $y$.

Equation (\ref{rec1}) describes stable and unstable solutions of
hamiltonian (\ref{ham}). In order to assess the stability
properties of the solutions, we need to study the hessian of the
system,
\begin{equation}
{\cal H}_N(y)=\left(
\begin{array}{ccccc}
d_1(y) & -1 & 0 & \ldots & 0 \\
-1  & d_2(y) & -1 & \ldots & 0 \\
0 & -1  & d_3(y) & \ldots & 0 \\
\vdots & \vdots & \vdots & \ddots & \vdots \\
0 & 0 & 0 & \ldots & d_N(y)
\end{array}
\right),
\end{equation}
where $d_n(y)=2+RV''(Y_n(y))$, in order to find out the stability
of solutions. Calling $\Delta_n(y)$ the determinant of the
principal minor of order $n$ of the hessian ${\cal H}_N(y)$, i.e.,
\begin{equation}
\Delta_n(y)=\det({\cal H}_n(y)),
\end{equation}
a stable solution must satisfy $\Delta_n(y)>0$ for all
$n=1,2,\ldots,N$. As the hessian is a tridiagonal matrix, there is
a recursive relation between different $\Delta_n$,
\begin{equation}\label{det}
\Delta_{n+1}(y)=d_{n+1}(y)\Delta_n(y)-\Delta_{n-1}(y),
\end{equation}
with $\Delta_1=d_1$ and $\Delta_2=d_1d_2-1$.

Expression (\ref{det}) above can be rewritten in terms of
$Y'_n(y)$. By deriving expression (\ref{rec1}) with respect to $y$
we find:
\begin{equation}\label{deriv}
Y'_{n+1}(y)=\frac{d\,Y_{n+1}(y)}{dy}=
\left[2+RV''(Y_n(y))\right]Y'_n(y)-Y'_{n-1}(y)=d_n(y)Y'_n(y)-Y'_{n-1}(y),
\end{equation}
with $Y'_2(y)=\Delta_1(y)$ and $Y'_3(y)=\Delta_2(y)$. Therefore, it has to be
\begin{equation}
\Delta_n(y)=Y'_{n+1}(y),
\end{equation}
for $n=1,2,\ldots,N$, and hence the stability region of solutions
(\ref{rec1}) are the points that satisfy $Y'_n(y)>0$ for all
$n=2,3, \ldots, N+1$.

This far, no approximations where needed to obtain these results,
still valid for any $V(Y)$. From now on, we will focus on the PB
model by choosing the Morse potential as our $V(Y)$, in order to
search for an analytic expression of the solutions (\ref{rec1}),
as well as to find the stability in terms of the initial condition
$y$.

\section{Continuum limit of the Peyrard-Bishop model}\label{sec:cont}

This limit corresponds to taking $R\ll 1$, which means that we can use the
approximation $Y_{n}(y)\rightarrow Y_{cont}(x,y)$ with
$x=n\sqrt{R}$ in Eq.\ (\ref{rec1}). By so doing we obtain
the following differential equation:
\begin{equation}\label{edo}
\frac{\partial^2Y_{cont}}{\partial x^2}=\frac{d\,V}{dY_{cont}},
\end{equation}
which can be easily solved using the initial conditions $Y_{cont}(x_0,y)=0$
and $\partial\,Y_{cont}(x_0,y)/\partial x=y/\sqrt{R}$,
where $x_0$ stands for the initial site of the model. This solution is
\begin{equation}\label{dw2}
e^{Y_{cont}(x,y)}=\frac{y\sqrt{\frac{2}{R}}\sinh\left[\sqrt{2+\frac{y^2}{R}}
(x-x_0)+\sinh^{-1}\left(\frac{y}{\sqrt{2R}}\right)\right]+2}{2+\frac{y^2}{R}}
\end{equation}
for $x\geq x_0$. Looking at expression (\ref{dw2}) we see that,
for a finite system, taking $y=0$ implies $Y_{cont}(x,0)=0$ for
all $x\geq x_0$. This result differs from the domain wall obtained
in the continuum limit in the PB model \cite{nonlinear} because in
this case we have restricted ourselves to finite systems ($x\geq
x_0$). For infinite systems, letting $Y_{cont}(x_0,y)\to 0$ as
$x_0\to\infty$, it can be seen that there is another stationary
solution, namely
\begin{equation}\label{dw1}
e^{Y(x)}=1+e^{\sqrt{2}x},
\end{equation}
In this respect, we believe that this solution should not be used
as an approximation of a finite system because it is not a
critical point of the continuum version of Eq.\ (\ref{ham}) and, 
therefore, we cannot speak
of stability in this case as long as we consider DNA as a finite
lattice.

\begin{figure}
\begin{center}
\includegraphics[height=6cm]{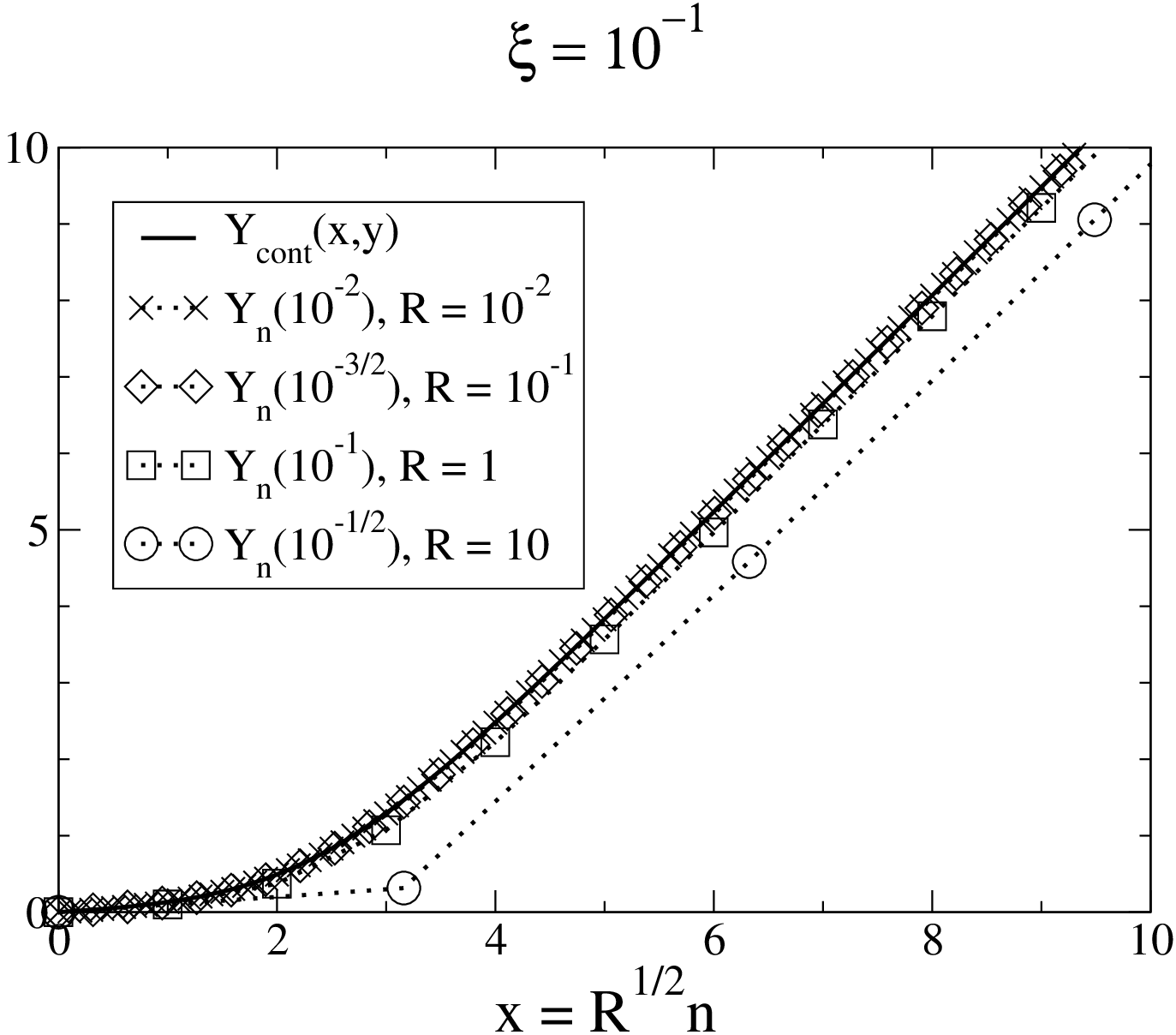}
\ \ \
\includegraphics[height=6cm]{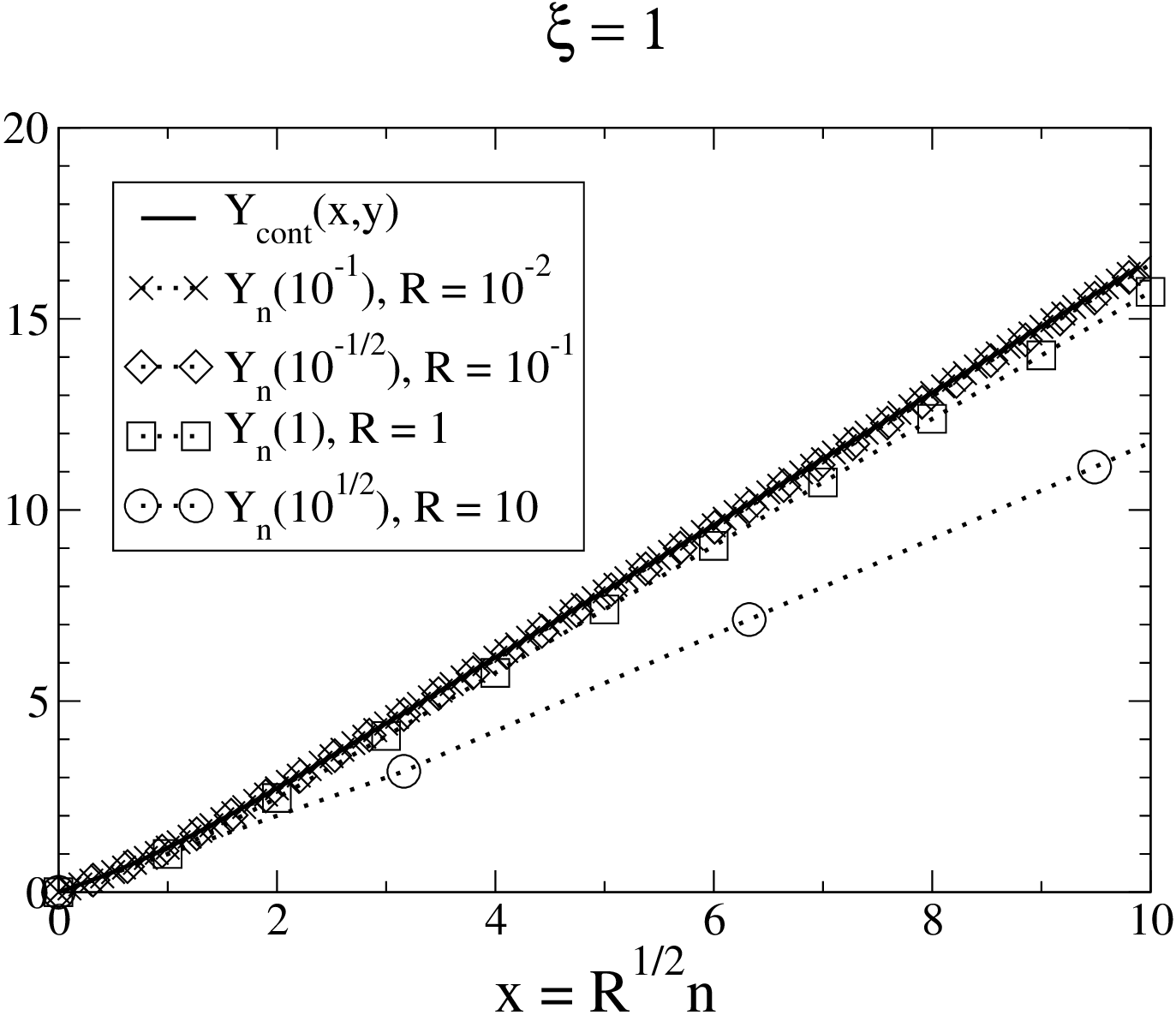}
\caption{\label{fig:cont} Plots of $Y_{cont}(x,y)$ (obtained from expression
(\ref{dw2})) for $\xi=0.1$ (left) and $\xi=1$ (right), where
$\xi=y/\sqrt{R}$, compared to the exact solutions $Y_n(y)$ of the
discrete recurrence relation (\ref{rec1}) calculated for the same
quotient $\xi=y/\sqrt{R}$ but for different values of $R$ (and,
therefore, different values of $y$). }
\end{center}
\end{figure}

An important feature of the continuum aproximation is that, in
Eq.\ (\ref{dw2}), $Y_{cont}$ can be written as a function of $x$
and $\xi$, where $\xi=y/\sqrt{R}$. This implies a scaling relation
between these two parameters, a relation that is absent for
solutions of the discrete limit. This behavior can be seen in
Fig.\ \ref{fig:cont}, where we represent $Y_{cont}$ with respect
to $x$ (with $x_0=0$) for two values of $\xi$, compared to the
exact result $Y_n(y)$ (recall that $x=n\sqrt{R}$) for different
values of $R$, and with $y=\xi\sqrt{R}$. We clearly observe that
for the largest values of $R$ ($R=1$ and, mostly, $R=10$) the
scaling relation is not fulfilled, indicating the crossover to the
discrete limit regime.

\begin{figure}
\begin{center}
\includegraphics[height=6cm]{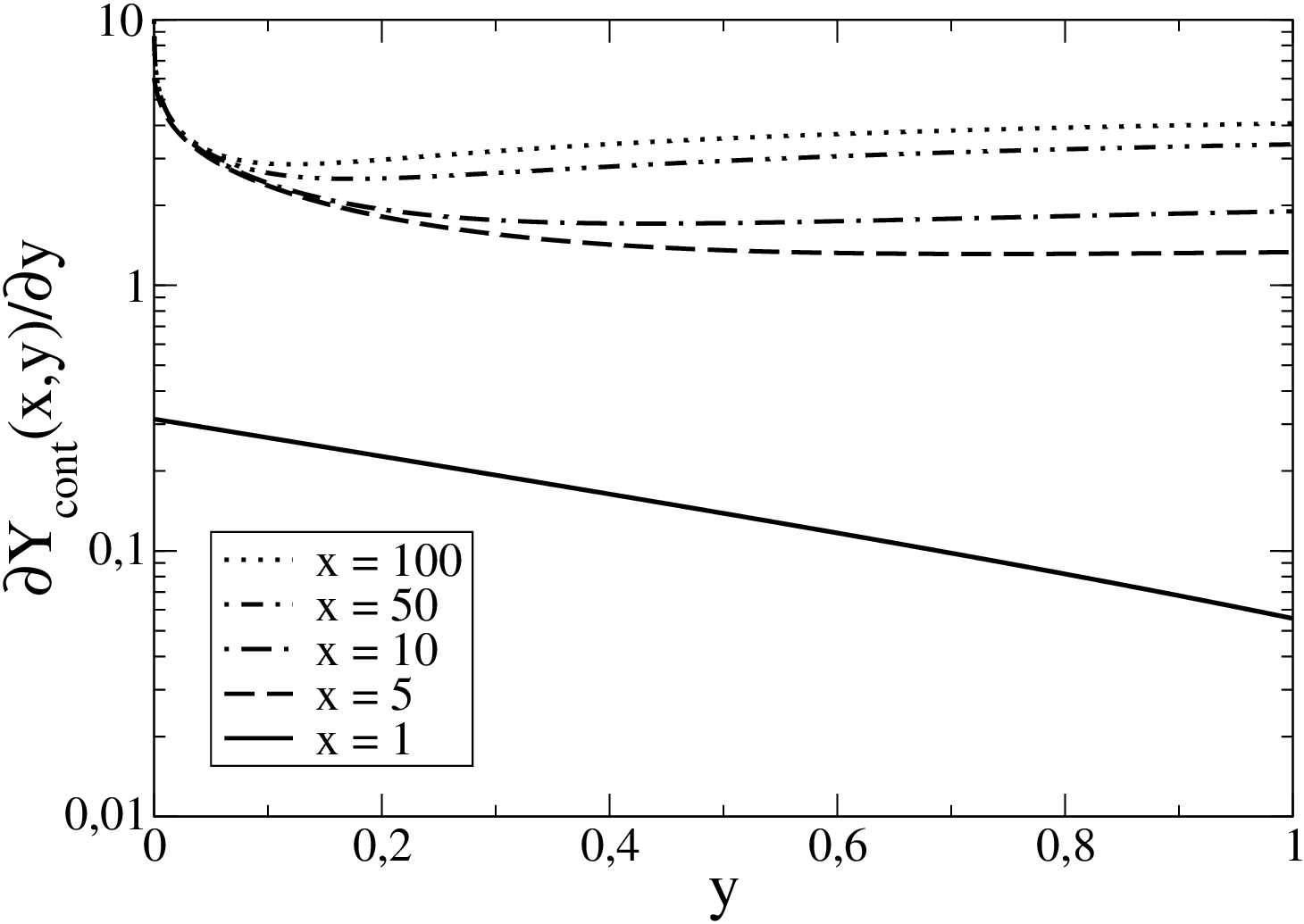}
\caption{\label{fig:stabcont} Plots of
$\partial\,Y_{cont}(x,y)/\partial y$ for (from lower to higher)
$x=1$, $x=5$, $x=10$, $x=50$ and $x=100$. All of them are positive
for any $y>0$, so all solutions of the form (\ref{dw2}) are stable
for any $y>0$.}
\end{center}
\end{figure}

To analize the stability of these solutions, using the result of
section \ref{sec:model}, it is enough to study the sign of
$\partial\,Y_{cont}(x,y)/\partial y$ for all $x\geq 0$. As the
derivative of expression (\ref{dw2}) is quite cumbersome, we
prefer to show plots of the result for different values of $x$,
which we collect in Fig. \ref{fig:stabcont}. It can be shown in
general that $\partial\,Y_{cont}(x,y)/\partial y>0$ for all $x>0$,
and therefore Eq.\ (\ref{dw2}) is a stable solution of
(\ref{rec1}).

In order to check these results, we used the equations of motion
of the model (\ref{ham}) in order to simulate the dynamics of
initial data given by (\ref{dw2}) with small perturbations, and
then monitored the evolution of this curves in time. Fixed
boundary conditions at both ends of the simulated interval were
used, in order to prevent the chain from spontaneously closing: We
note that the global minimum of the hamiltonian (\ref{ham}) is the
null solution. Therefore, in order to verify our results we had to
restrict the simulations to the sector of open-chain solutions by
choosing those boundary conditions. With that caveat, our
simulations fully confirm the predicted stability of solutions. We
stress that such solutions are the ones that are relevant to the
mechanical denaturation problem, where the spontaneous closing of
the chain is prevented by the force exerted on the open end.

\section{Discrete limit of the Peyrard-Bishop model}\label{sec:discrete}

\subsection{Solutions}

\begin{figure}
\begin{center}
\includegraphics[height=6cm]{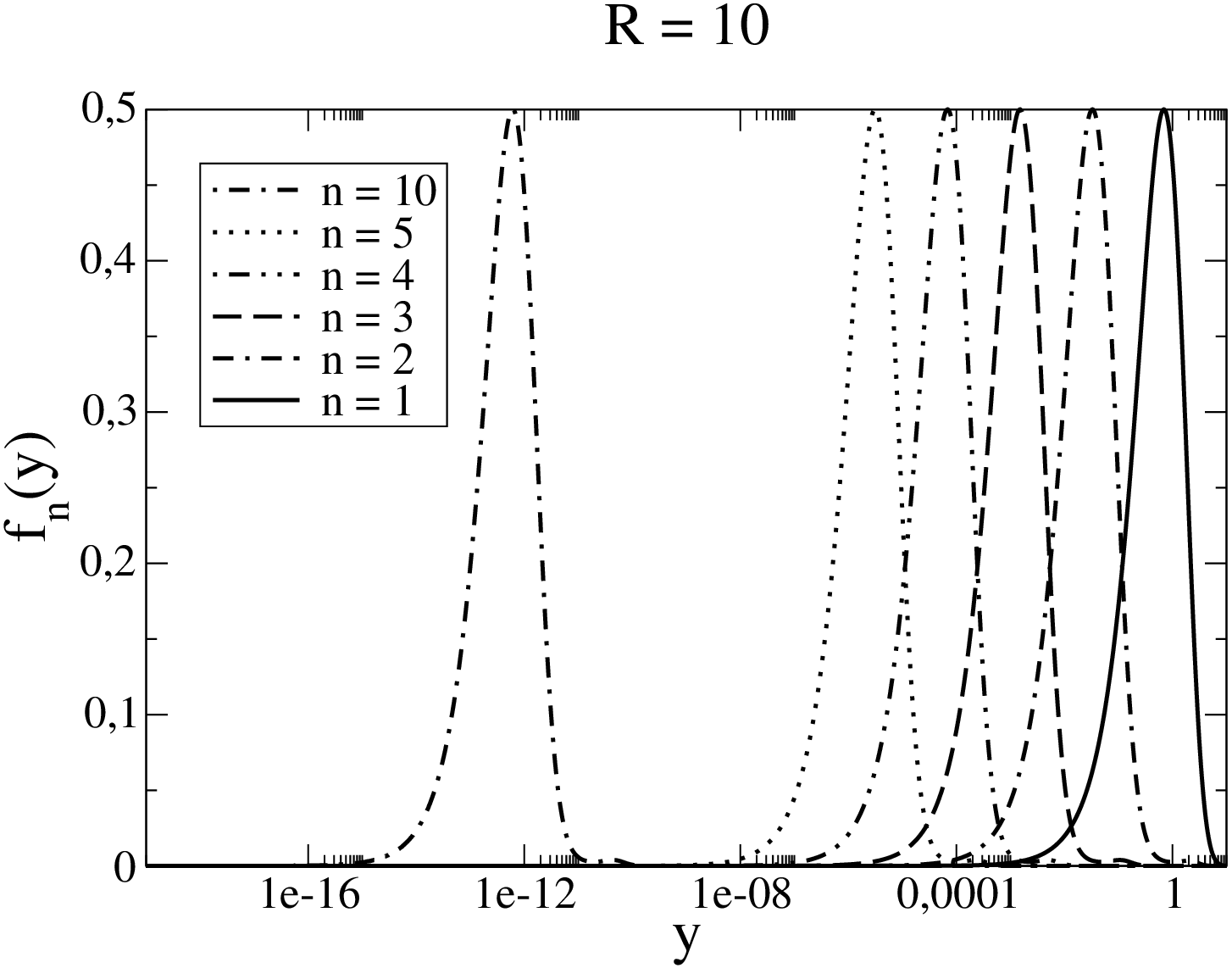}
\includegraphics[height=6cm]{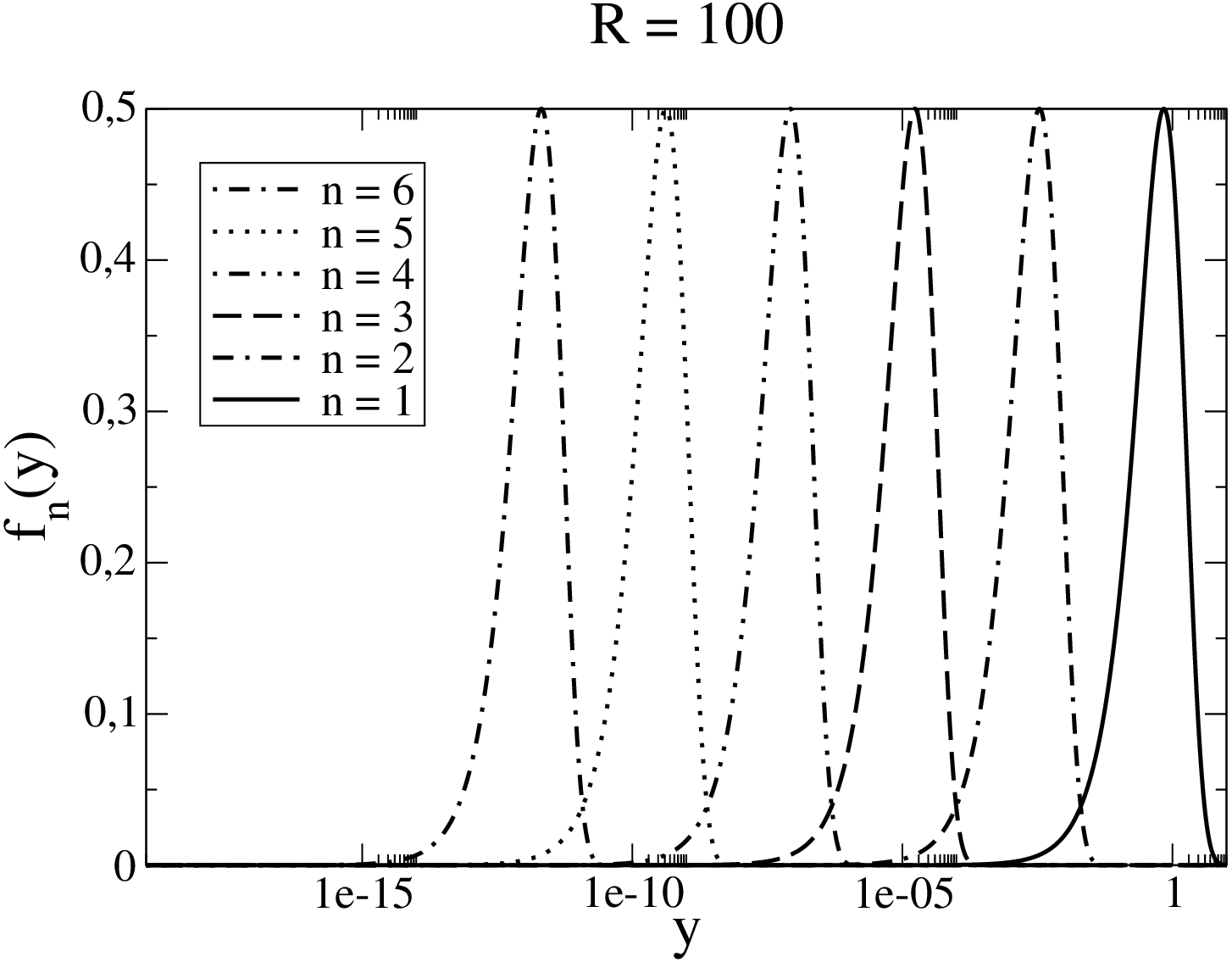}
\caption{\label{fig:fnum} Functions $f_n(y)$ for $R=10$ (left) and
$R=100$ (right), for different values of $n$. }
\end{center}
\end{figure}

\begin{figure}
\begin{center}
\includegraphics[height=6cm]{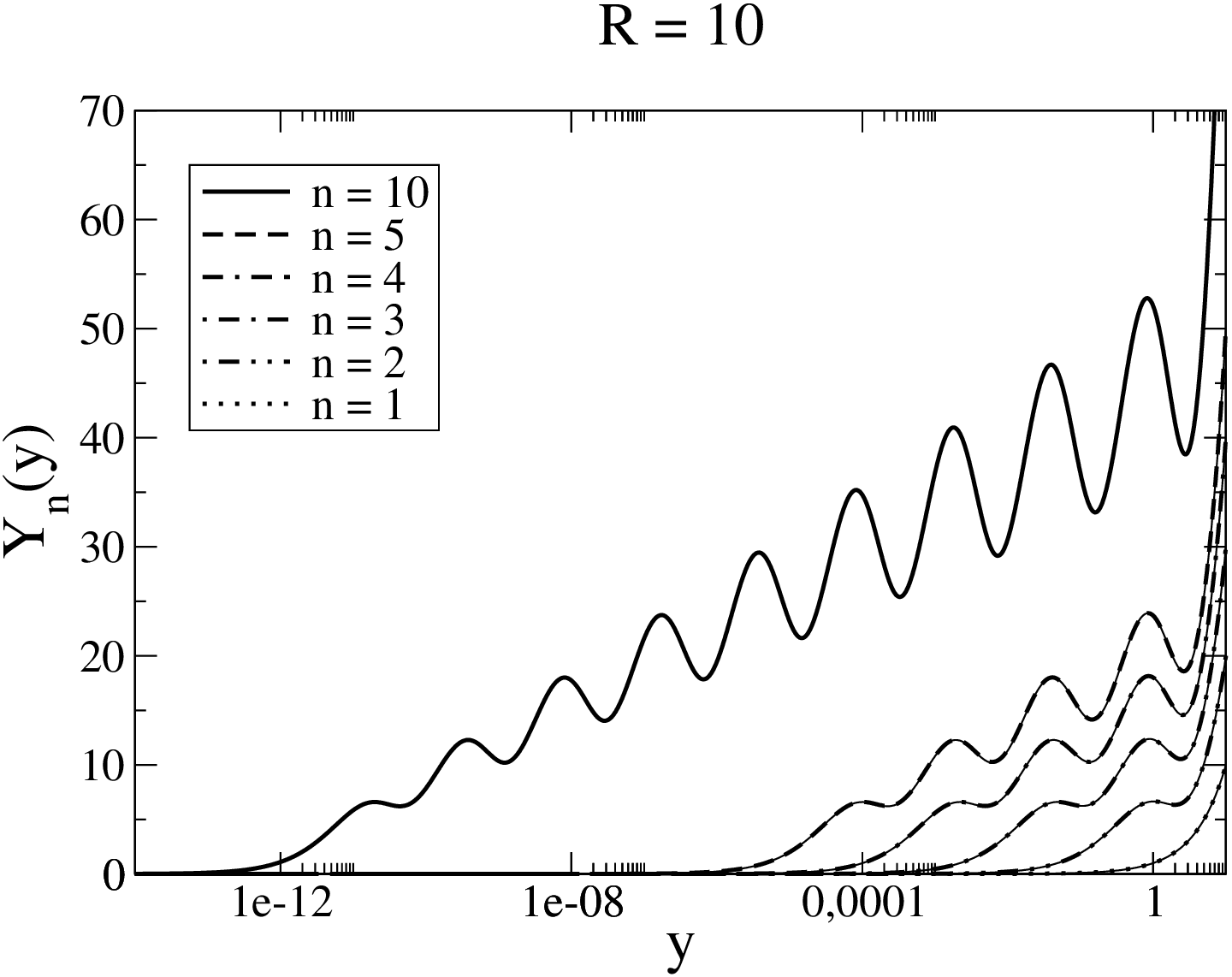}
\includegraphics[height=6cm]{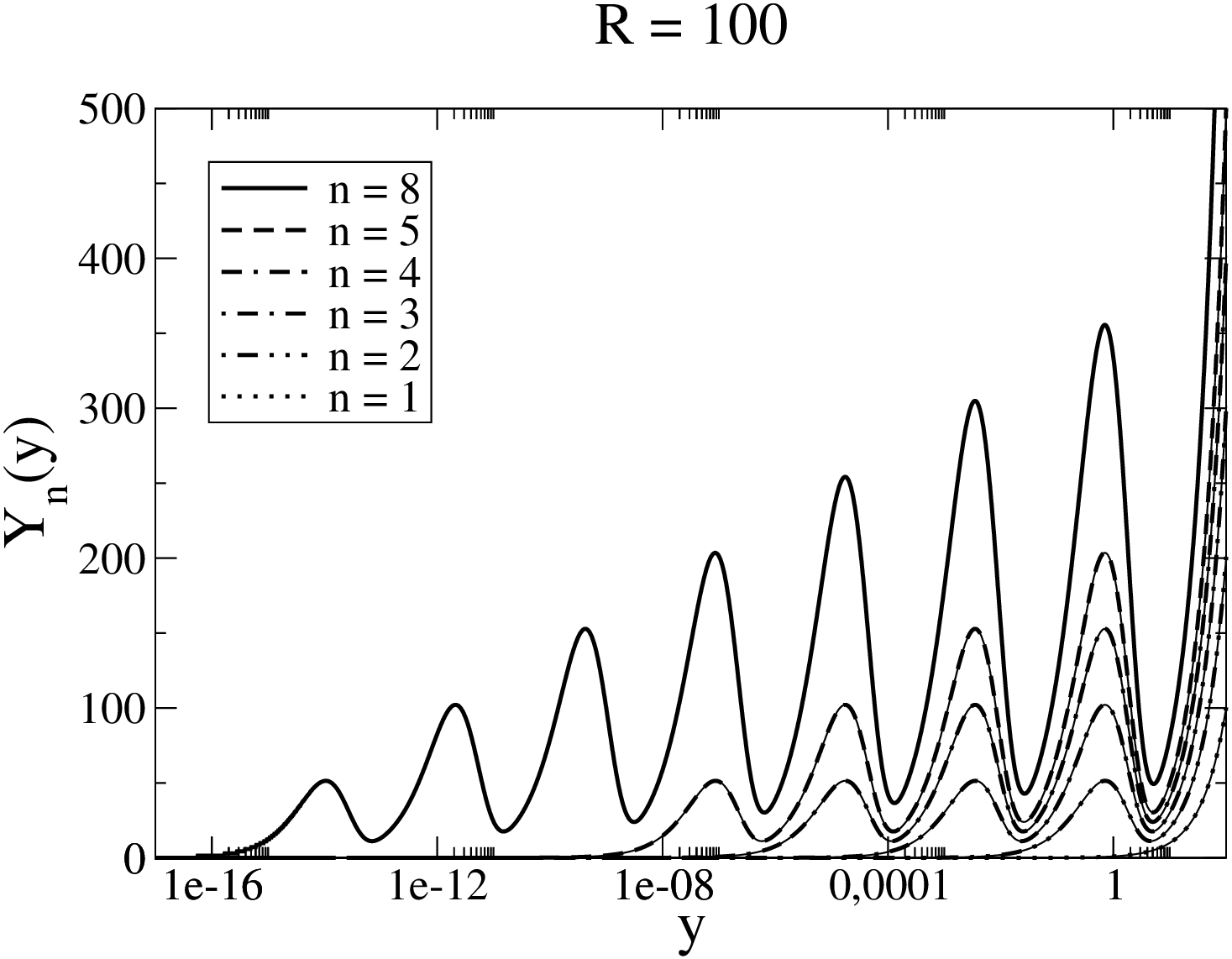}
\caption{\label{fig:ap2} Approximation $Y^{(2)}_n(y)$ (see Eq.\
(\ref{ap2})) vs. exact $Y_n(y)$, for $R=10$ (left) and $R=100$
(right), for different values of $n$. Exact solutions are drawn
with thick lines, whereas the corresponding approximations are
drawn in thin, solid lines. }
\end{center}
\end{figure}

The discrete limit of the PB model corresponds to letting $R\gg
1$, and can be obtained following a few steps. Using a telescopic
summation of $Y_{n+1}-Y_n$, and noting the initial conditions
$Y_0=0$ and $Y_1=y$, Eq.\ (\ref{rec1}) can be rewritten as
\begin{equation}\label{rec2}
Y_{n+1}(y)=(n+1)y+R\sum_{k=1}^n(n+1-k)\,V'(Y_k(y)).
\end{equation}
We now define
\begin{equation}
f_k(y)\equiv
V'(Y_k(y))=2e^{-Y_k(y)}\left(1-e^{-Y_k(y)}\right)=f_1(Y_k(y)).
\end{equation}
These functions are plotted for different values of $R$ in the
discrete limit in Fig. \ref{fig:fnum}.
As can be seen, these $f_k$ are very localized, their overlapping
depending on $R$. In fact, in the discrete limit we are working
on, which implies low overlapping of the curves, we can calculate
the position of the maxima of each $f_k(y)$, and subsequently
approximate $f_k(y)$ by the first function, $f_1(y)$, by
writing
\begin{equation}\label{fap1}
f_k(y)\simeq f^{(1)}_k(y)\equiv f_1(b_ky),
\end{equation}
with
\begin{equation}
b_n=\frac{1}{2\sqrt{R(R+2)}}
\left[\left(R+1+\sqrt{R(R+2)}\right)^n-\left(R+1-\sqrt{R(R+2)}\right)^n\right].
\end{equation}
This result allows us to obtain an analytic, approximate
expression of the solutions for different $y$ in the discrete
limit. Substituting it in Eq.\ (\ref{rec2}), we find that
\begin{equation}\label{ap1}
Y^{(1)}_{n+1}(y)=(n+1)y+R\sum_{k=1}^n(n+1-k)\,f^{(1)}_k(y).
\end{equation}
is a good approximation of the exact solution $Y_n(y)$ for large
values of $R$.
\begin{figure}
\begin{center}
\includegraphics[height=6cm]{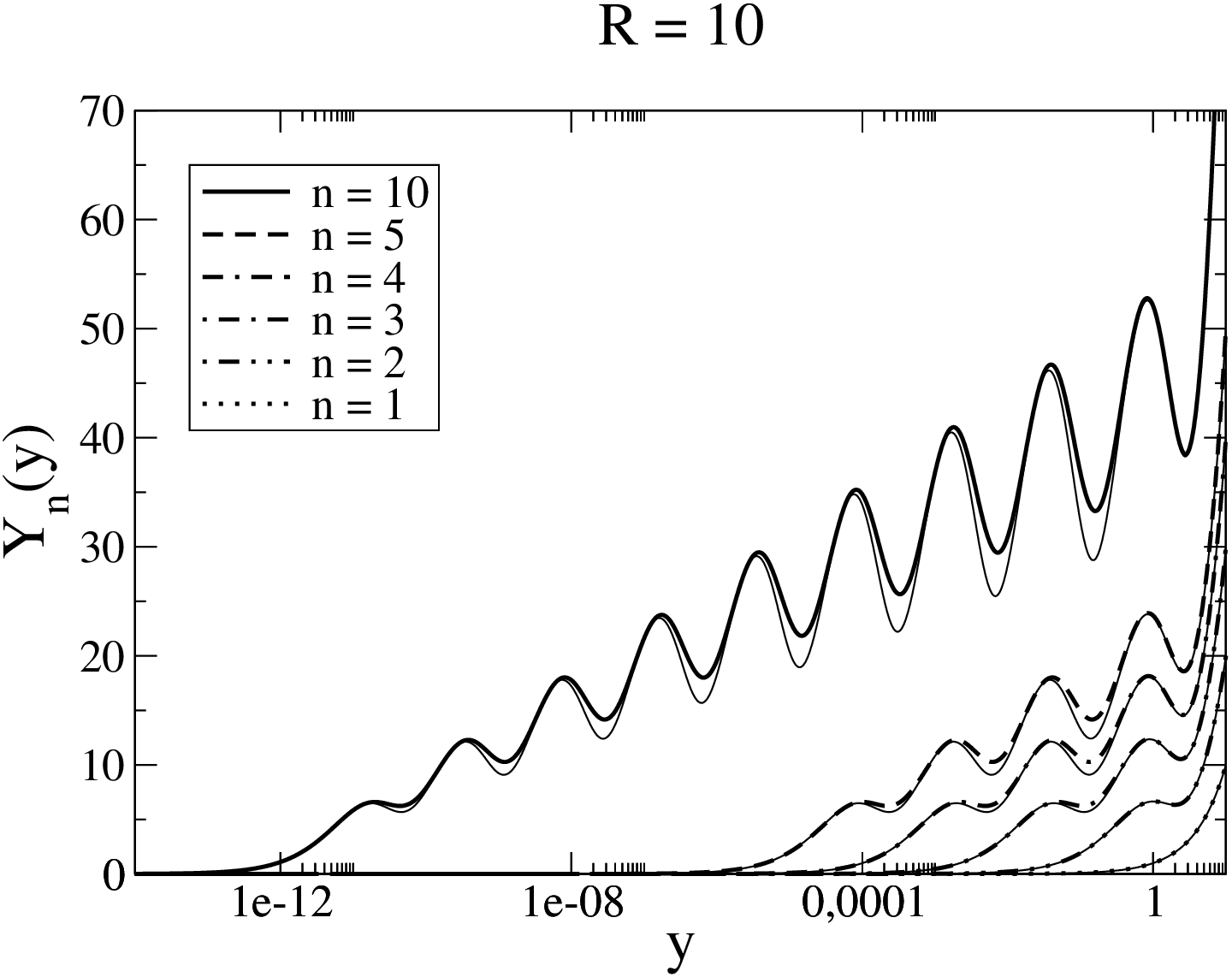}
\includegraphics[height=6cm]{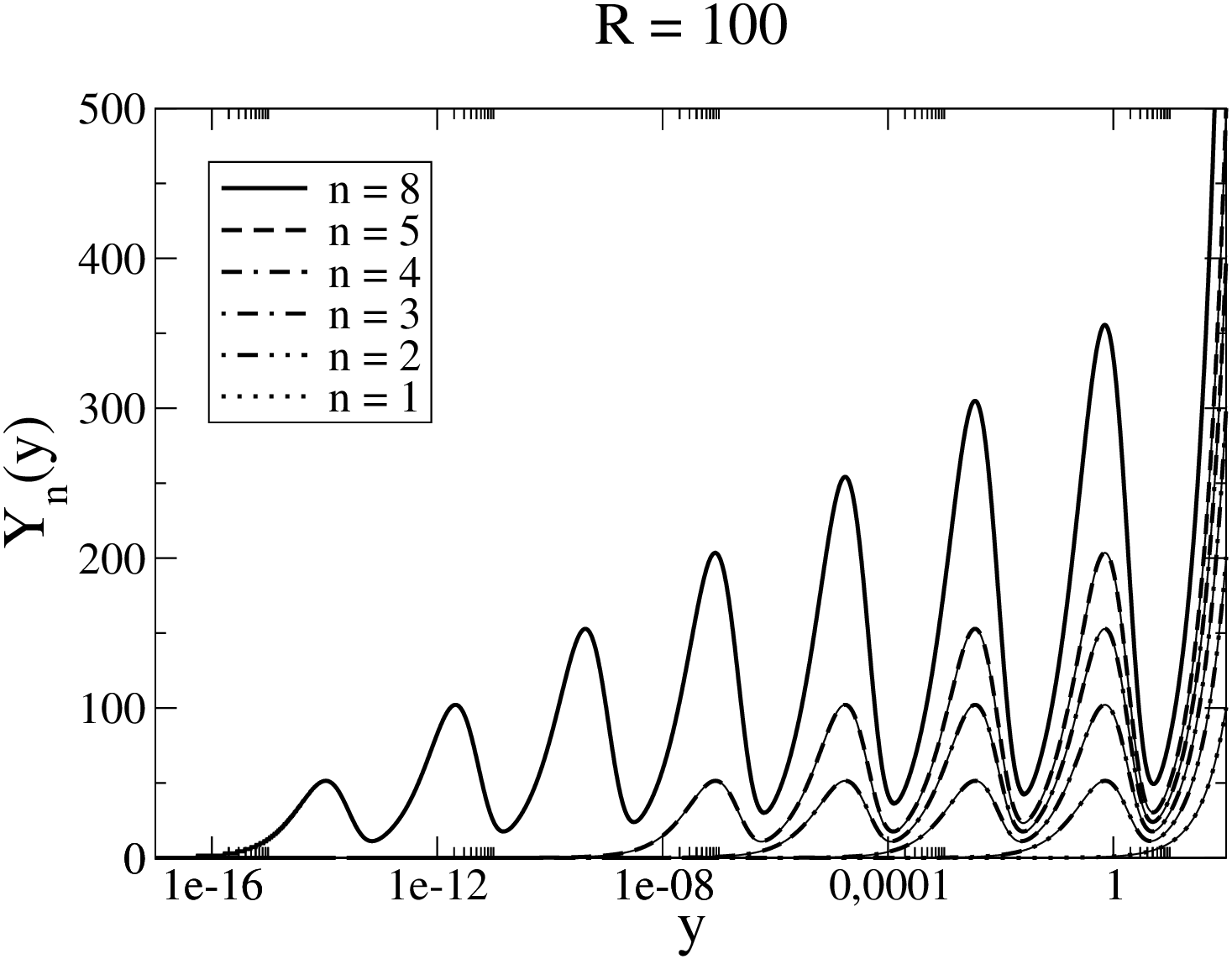}
\caption{\label{fig:ap1} Approximation $Y^{(1)}_n(y)$ (see Eq.\
(\ref{ap1})) vs. exact $Y_n(y)$, for $R=10$ (left) and $R=100$
(right), for different values of $n$. Exact solutions are drawn
with thick lines, whereas the corresponding approximations are
drawn in thin, solid lines. }
\end{center}
\end{figure}
Figure \ref{fig:ap1} confirms the accuracy of this approximation:
for $R\gtrsim 100$, the approximation is very accurate, whereas
the smaller $R$ the worse the approximation. For smaller values of
$R$, the approximation can be improved by resorting to the next
function, $f_2(y)$ instead of $f_1(y)$, as a substitute for the
rest of the $f_k(y)$, by defining
\begin{equation}\label{fap2}
f_k(y)\simeq f^{(2)}_k(y)\equiv f_2(y\frac{b_k}{b_2})
\end{equation}
for $k=3,4\ldots$, and approximating $Y_n(y)$ by
\begin{equation}\label{ap2}
Y^{(2)}_{n+1}(y)= (n+1)y+nRf_1(y)+R\sum_{k=2}^n(n+1-k)\,f^{(2)}_k(y).
\end{equation}
In this case, the approximation is even better than for $Y^{(1)}_n(y)$,
and even for $R=10$ the results are very close to the exact ones
(see Fig. \ref{fig:ap2} for details).

The errors of these approximations depend on the value of $R$ and
$n$. For instance, $Y^{(1)}_n(y)$ is exact for $n=1$ and $n=2$,
for any value of $R$, whereas $Y^{(2)}_n(y)$ is exact up to $n=3$
for any $R$. For low values of $n$, the main difference between
the exact $f_n(y)$ (which can be easily obtained numerically) and
$f^{(1)}_n$ is located around the maxima of $f_n(y)$, with a
maximum error $E^{(1)}_{max}\simeq 0.06$ for $R=10$ and
$E^{(1)}_{max}\simeq 0.006$ for $R=100$. For the second order
approximation based on $f^{(2)}_n$, the maximum error is
$E^{(2)}_{max}\simeq 2.7\, 10^{-3}$ for $R=10$ and
$E^{(2)}_{max}\simeq 3.2\, 10^{-5}$ for $R=100$; at the same time,
another discrepant region, much less so than the main one, appears
around the position of the maxima of $f_{n-1}(y)$. The same
calculation can be done for higher orders of the approximating
function, $f^{(k)}_n(y)$, and it can be seen that the reduction of
the error using $k+1$ instead of $k$ is at least of one order of
magnitude. There is a computational limit near the precision of
the machine, which does not allow us to check the validity of this
assumption further than a certain $k$ and $n$, depending of the
value of $R$, but, as far as we know, it is reasonable to expect
that the same behavior will take place for higher values of $k$
and $n$. Therefore, we conjecture that higher orders of functions
$f_k(y)$ can be used as approximations of $f_{n}(y)$, as
$f_n(y)\simeq f^{(k)}_n(y)$, with
\begin{equation}
f^{(k)}_{n}(y)= f_k\left(y\frac{b(k')}{b(k)}\right)
\end{equation}
for $n>k$, in order to obtain
better approximations $Y^{(k)}_n(y)$ of the exact solution $Y_n(y)$,
and that the error of an approximation of order $k$, $E^{(k)}_n(y)=
Y_n(y)-Y^{(k)}_n(y)$, can be estimated as the difference
\begin{equation}
E^{(k)}_n(y)\simeq Y^{(k+1)}_n(y)-Y^{(k)}_n(y)+
\mathcal{O}\left(E^{(k+1)}_n(y)\right),
\end{equation}
with $E^{(k+1)}_n(y)\ll E^{(k)}_n(y)$.

\subsection{Stability}

The approximations defined in (\ref{ap1}) and (\ref{ap2}), as well
as the ones mentioned in the above section allow us to calculate
very accurately the solution $Y_n(y)$ for any value of $n$ and
$y$. This is important because in the exact, numerical calculation
of Eqs.\ (\ref{rec1}) and (\ref{rec2}) there are problems for
values of $y$ close to zero, due to the numerical precision of the
computer (see also the discussion below). Therefore, for analyzing
the stability in the discrete limit, we proceed to use the
approximations $Y^{(k)}_n(y)$ previously discussed. By this means,
we can work with systems of much larger size than the ones that
could be studied solving numerically the original recurrence
relations.
\begin{figure}
\begin{center}
\includegraphics[height=6cm]{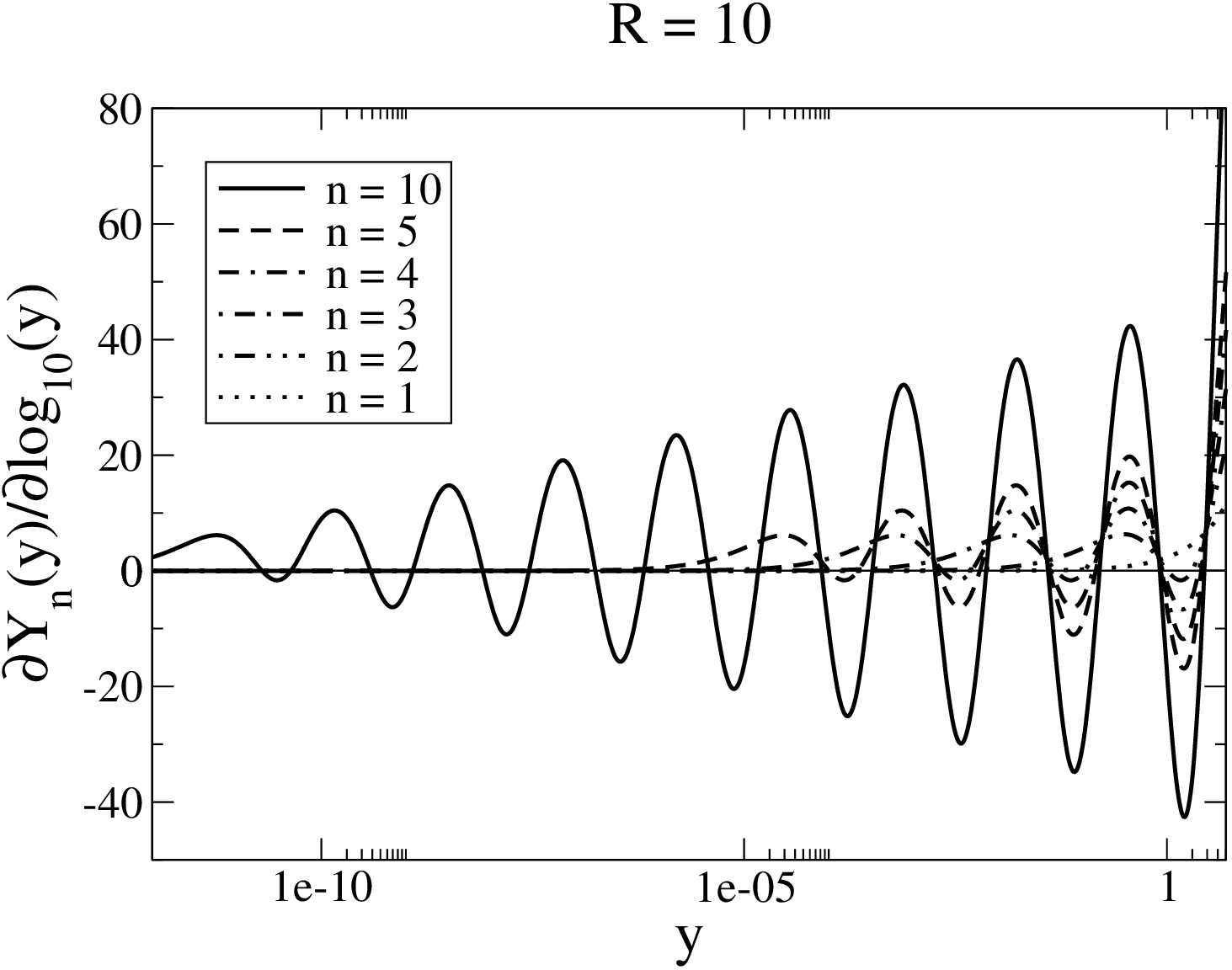}
\includegraphics[height=6cm]{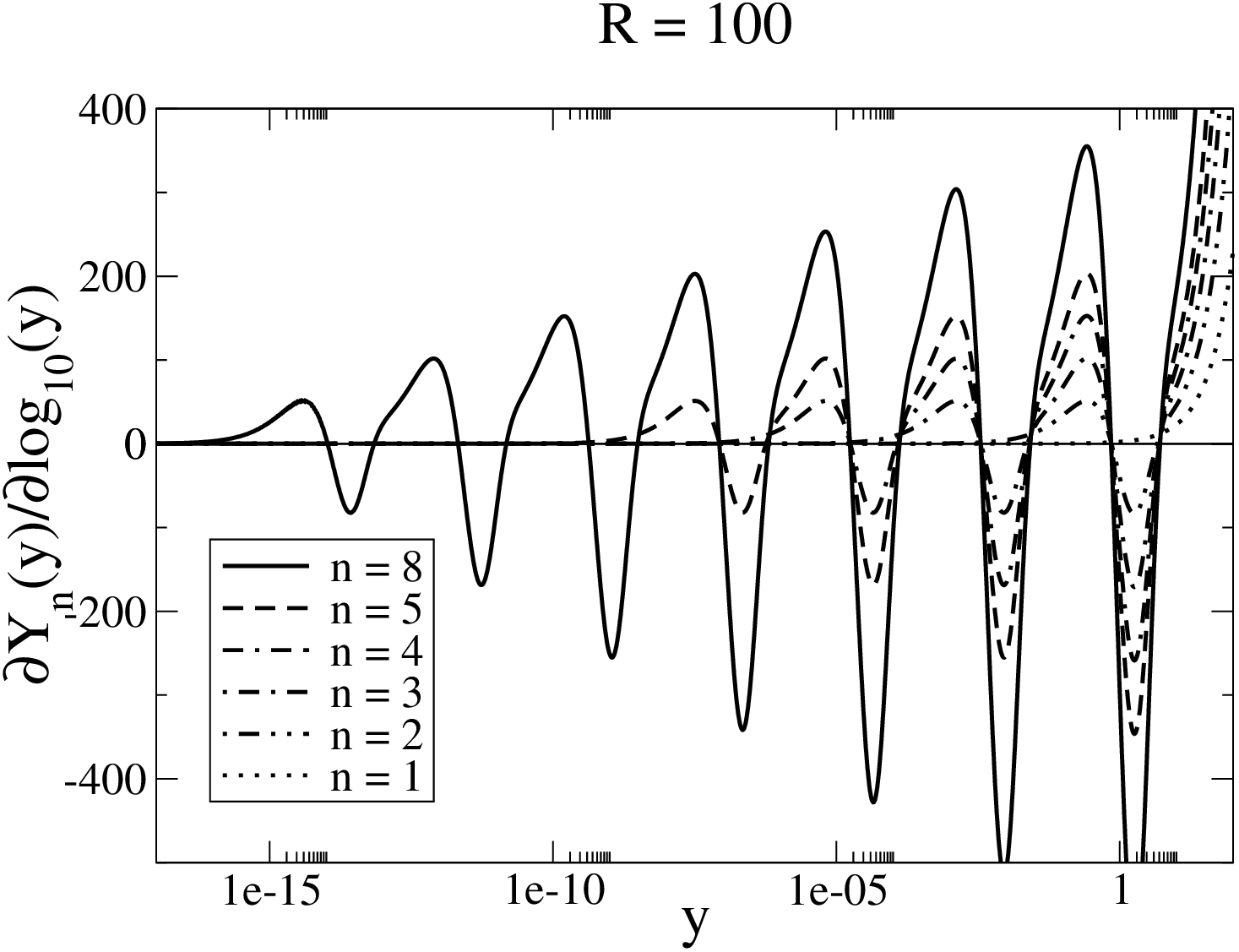}
\caption{\label{fig:stab} Derivative of $Y_n(y)$ with respect to
$log_{10}(y)$ for different values of $n$ and for $R=10$ (left)
and $R=100$ (right), with logarithmic $x$ axis. The stability
region of a system of size $N$ is the intersection of all the
points that satisfy $\partial\, Y/\partial y>0$ for
$n=1,2,\ldots,N+1$. From the figures, we find that the stability
region corresponds to the points that satisfy the condition for
$n=N+1$, as the stability region of a system of size $N$ seems to
be embedded in the stability region of a system of size $N-1$ (see
text for an explanation). }
\end{center}
\end{figure}
For comparison, in the study of stability we will show results for
systems of small size, where the derivative $Y'_n(y)$ can be
calculated without approximations for each $n$ without high errors
of the precision of the computer. In Fig.\ \ref{fig:stab} we show
the dependence of $\partial Y_n(y)/\partial \log_{10}(y)$ on
function of the initial condition $y$, in logarithmic scale, for
different values of the size of the system, $N$. We chose
$\partial Y_n(y)/\partial \log_{10}(y)$ instead of $Y'_n(y)$ in
order to obtain a smooth curve: The direct plot of $Y'_n(y)$ would
make very difficult to observe the intervals with $Y'_n(y)>0$. As
the sign of both derivatives is the same for all $y>0$, we have
resorted the logarithmic one. With this change, a modulated
''sinusoidal'' structure reveals itself in Fig. \ref{fig:stab} for
each $n$, with $n-1$ maxima and minima around $Y'_n(y)=0$. From
that figure, it is apparent that a new interval of instability for
lower values of $y$ appears in systems of size $n$ as compared to
systems of size $n-1$. In addition to this, Fig.\ \ref{fig:stab}
also suggests that the set of unstable points of a system of size
$n$ containes the set of unstable points of a system of size
$n-1$. A plausibility argument for this statement goes as follows:
Let us look at points that satisfy $Y'_n(y_0)=0$, in the extremes
of an interval of unstable points. Then, if $Y'_{n-1}(y_0)>0$, it
must be $Y'_{n+1}(y_0)<0$ (see Eq.\ (\ref{deriv})), and therefore
the interval of unstable points for a system of size $n$ will be
larger than for a system of size $n-1$. This condition is
satisfied by all the new unstable intervals that appear for each
$Y_n(y)$, starting on $Y_2(y)$, and therefore, by induction, it
can be applied to all systems. Therefore, all stable points of a
system of size $n$ are those who satisfy $Y'_{n+1}(y)>0$.

As an independent check of the validity of the results shown in
this section, we compared our results with the ones recently
reported in \cite{pbdis}. By studying the discrete, stationary
problem with fixed boundary conditions, they found eight stable
and seven unstable solutions of a system of size $N=28$. The
specific boundary conditions they used were $Y_0=0$ and $Y_N=80$
for $R=10.1$.
\begin{figure}
\begin{center}
\includegraphics[height=6cm]{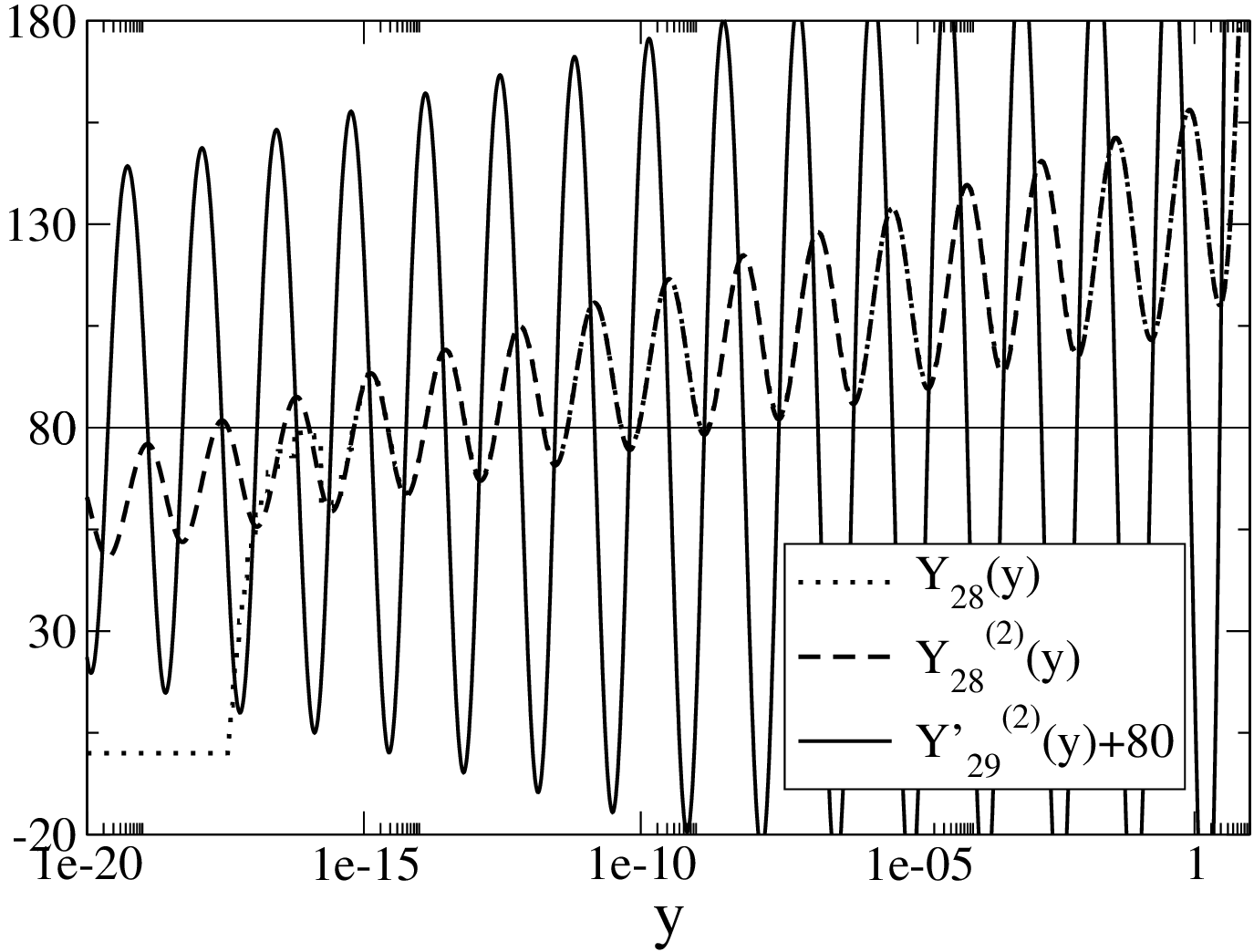}
\caption{\label{fig:comp} Stable and unstable solutions given by
approximation $Y_N^{(2)}$ for a system with $N=28$ sites, $Y_0=0$
and $Y_N=80$. The solutions are the intersections of the curves
with the line $Y_{28}=80$. It is necessary to study the sign of
$Y'_{29}$ on each solution in order to establish the stability of
the solutions. Once again, $\partial Y_{29}(y)/\partial
\log_{10}(y)$ is plotted for clarity. Our approximation gives
eight stable and seven unstable solutions, exactly as in
\cite{pbdis}. }
\end{center}
\end{figure}
The exact $Y_n(y)$ and the approximate solution $Y^{(2)}_n(y)$ of
that system are in Fig. \ref{fig:comp}. The plot now makes clear
the precision problem we mentioned above, namely when we tried to
calculate the exact, numerical solution for low values of $y$. On
the other hand, the approximate solution $Y^{(2)}_n(y)$ was
calculated without any problem in a wide range of $y$. It is also
shown that $Y^{(2)}_n(y)$ gives the same number of both stable and
unstable solutions as in \cite{pbdis} (see explanation in
caption), which implies that the structure of peaks of $Y_n(y)$
gives a good explanation of the number and structure of solutions.
We think that this method can be applied for larger systems with
the way to estimate errors that we explained in this section.

\section{Conclusions}\label{sec:conclusions}

In this paper, we have reported a study of the stationary
solutions of the PB model, obtaining exact and analytical
approximations of the continuum and the discrete limit. We have
been able to obtain all the stationary solutions and to classify
them according to their stability by considering the stationary
equation as an initial value problem. We have also found that, in
the discrete limit, the exact solutions can be approximated to the
desired degree of accuracy by using the functions $f_k(y)$ as
explained above. We have compared our results obtained in the
discrete limit with \cite{pbdis} finding a very good agreement
with the number of stable and unstable solutions of a PB system
with fixed boundary conditions, thus giving an explanation of the
multiple solutions of the problem and the stability. In fact, our
results show that every solution of the initial value problem,
which is unique for every choice of $y$, corresponds to exactly
one of the problem with fixed boundary conditions \cite{pbdis},
which does not have a unique solution. This is the reason why the
picture we are providing here is much more comprehensive and
allows to understand fully the space of solutions of the problem.
On the other hand, the method explained in this paper to obtain
stable solutions of a system of size $N$ and opening $L$
allows to work with larger systems, as solutions and their stability
are calculated by evaluating a function, instead of numerically
(as in \cite{pbdis}, with $N=28$). 
We also believe that this study may be extended to the more accurate
description of DNA given by the Peyrard-Bishop-Dauxois model
\cite{DPB}, where the coupling between two consecutive bases of
the DNA molecule has an anharmonic term that affects the general
behavior of the openings \cite{campa,LosAlamos}. 

As stated in the introduction, this stems from previous studies in
the sine-Gordon (Englander) model of DNA \cite{sara1, sara2},
where the relation between the dynamics of soliton-like
excitations and the inhomogeneity of the DNA sequence was studied.
In fact, what we are reporting here is only the first step towards
the study of an effective potential that may explain the dynamics
of these stationary, stable solutions in presence of
heterogeneities in the sequence and an external force. Once an
analytical expression of the stationary solutions of the model is
found, as we have just done, we will resort to a collective
coordinate technique to find an {\em effective potential}
description of the dynamics. The final aim of such a program is to
find out whether this approach allows to identify important sites
from the genomic viewpoint along any given sequence. While work
along these lines is in progress, we believe that the richness of
the structure of the stationary solutions we have found and their
stability is of interest in itself and can motivate further
research in these and related models. Finally, we believe our
solutions can be exploited to analyze the statistical mechanics
of the PB model along the lines of \cite{pbdis,qasmi}, in
particular because of the advantage of having an approximate,
analytical expression.

\section*{Acknowledgments}

We thank Michel Peyrard and Fernando Falo for many discussions
about DNA and domain walls. This work has been supported by the
Ministerio de Educaci\'on y Ciencia of Spain through grants
BFM2003-07749-C05-01, FIS2004-01001 and NAN2004-09087-C03-03. S.C.
is supported by a fellowship from the Consejer\'\i a de
Edu\-ca\-ci\'on de la Comunidad Aut\'onoma de Madrid and the Fondo
Social Europeo.


\end{document}